\documentclass[letterpaper,aps,prd,twocolumn,floats,showpacs,preprintnumbers,nofootinbib,superscriptaddress]{revtex4}
\usepackage{graphicx}
\usepackage[intlimits,centertags]{amsmath}
\usepackage{amssymb,amsfonts}
\usepackage{gensymb}
\usepackage{mathrsfs}
\usepackage[dvips]{hyperref}

\medskipamount
\newcommand{\D}{\mathrm{d}}

\begin{document}

\preprint{OUTP-0904P}
\title{Neutrino diagnostics of ultra-high energy cosmic ray protons} 
\author{Markus Ahlers}
\affiliation{Rudolf Peierls Centre for Theoretical Physics, University of Oxford, Oxford OX1 3NP, UK\\\vspace{-0.1ex}}
\author{Luis A.~Anchordoqui}
\affiliation{Department of Physics, University of Wisconsin-Milwaukee, Milwaukee, WI 53201, USA}
\author{Subir Sarkar}
\affiliation{Rudolf Peierls Centre for Theoretical Physics, University of Oxford, Oxford OX1 3NP, UK\\\vspace{-0.1ex}}

\begin{abstract}
  The energy at which cosmic rays from extra-galactic sources begin to
  dominate over those from galactic sources is an important open
  question in astroparticle physics. A natural candidate is the
  energy at the `ankle' in the approximately power-law energy spectrum which is
  indicative of a cross-over from a falling galactic component to a
  flatter extra-galactic component. The transition can occur without
  such flattening but this requires some degree of conspiracy of the
  spectral shapes and normalizations of the two
  components. Nevertheless it has been argued that extra-galactic
  sources of cosmic ray protons which undergo interactions on the CMB
  can reproduce the energy spectrum below the ankle if the cross-over
  energy is as low as the `second knee' in the spectrum. This low
  cross-over model is constrained by direct measurements by the Pierre
  Auger Observatory which indicate a heavier composition at these
  energies. We demonstrate that upper limits on the cosmic diffuse
  neutrino flux provide a complementary constraint on the proton
  fraction in ultra-high energy extra-galactic cosmic rays and
  forthcoming data from IceCube will provide a definitive test of this
  model.
\end{abstract}


\maketitle

\section{Introduction}

The flux of cosmic rays (CRs) falls as an approximate power-law in
energy, $\mathrm{d}N/\mathrm{d}E \propto E^{-\gamma}$, with $\gamma
\simeq 2.8$ from about 1~GeV up to the `knee' in the spectrum at $\sim
3 \times 10^{6}$~GeV where it steepens to $\gamma \simeq 3$; it then
steepens further to $\gamma \simeq 3.2$ at $\sim 5 \times10^8$~GeV
(the `second knee') and flattens back to $\gamma \simeq 2.8$ at the
`ankle' at $\sim 3
\times10^9$~GeV~\cite{Nagano:2000ve,Amsler:2008zzb}. A long-standing
open question is the transition point between dominance by galactic
and extra-galactic sources in the spectrum of ultra-high energy (UHE)
CRs. The transition between the two components ought to be accompanied
by the appearance of spectral features, {\it e.g.}~two power-law
contributions would naturally produce a {\em flattening} in the
spectrum if the harder component dominates at lower energies. Hence,
the ankle seems to be a natural candidate for this
transition~\cite{Linsley:1963bk,Hill:1983mk,Wibig:2004ye} (for discussions see
Refs.~\cite{Hillas:2004nn,Stanev:2006ri}).

It has been argued however that the cross-over of the two
contributions might also happen at lower energies without a
corresponding flattening of the spectrum or even at a spectral {\em
  steepening} like the second knee~\cite{Berezinsky:2002nc}. Such a
transition would appear to require considerable fine-tuning of the
shape and normalization of the two spectra. It has been argued however
that the spectral shape required for the extra-galactic component can
develop naturally during the propagation of extra-galactic protons in
the cosmic microwave background (CMB) over cosmological distances. At
an energy of $\sim 3\times10^8$~GeV the energy losses due to $e^+e^-$
pair production and cosmic expansion are roughly equal and thus
produce a steepening of an initially featureless power-law injection
spectrum. In this model the ankle is formed as a result of the dip due
to $e^+e^-$ pair production together with a pile-up of protons below
the Greisen-Zatsepin-Kuzmin (GZK) suppression which sets in at about
$~6 \times10^{10}$~GeV due to pion
photoproduction~\cite{Greisen:1966jv,Zatsepin:1966jv}.

An obvious test of this `dip-transition'
model~\cite{Berezinsky:2002nc} is whether the flux is indeed proton
dominated at energies above $10^9$~GeV~\cite{Aloisio:2007rc}. Recent
measurements of the elongation rate of UHE CR air showers by the
Pierre Auger Observatory~\cite{Abraham:2004dt} indicate however
an increasingly {\em heavier} composition in the energy range
\mbox{$\sim 2 \times 10^9 - 4 \times
  10^{10}$~GeV}~\cite{Unger:2007mc}, in conflict with this
expectation. More elaborate
models~\cite{Allard:2005ha,Allard:2005cx,Hooper:2006tn} in which the
the extra-galactic CRs have a {\em mixed} composition and the
cross-over occurs between the second knee and the ankle are consistent
with the data~\cite{Allard:2007gx,Anchordoqui:2007fi}, but lack the
simplicity of the dip-transition model. However, this illustrates that
even precision measurements of the chemical composition and spectrum
of CRs may not be enough to resolve conclusively just where the
galactic-to-extra-galactic transition occurs.

An important clue in resolving this puzzle would be the identification
of the CR accelerators themselves. Indeed, the Auger data show a
correlation \mbox{(at $>99\%$ C.L.)} between the arrival directions of CRs
with energy above $\sim 6 \times 10^{10}$~GeV and active galactic
nuclei within \mbox{$\sim 75$~Mpc}~\cite{Cronin:2007zz,Abraham:2007si}. 
However, this has only deepened the puzzle, indicating the possible return 
of a {\em lighter} composition beyond $4\times10^{10}$~GeV, since the 
primaries appear not to have been deflected significantly by intervening 
inter-galactic and galactic magnetic fields and therefore are more likely 
to be protons rather than heavy nuclei. The cosmic ray sources ought also to be emitting other
messenger particles that are produced during the acceleration process,
in particular gamma rays and neutrinos, and these might provide
further constraints. Indeed, it has been observed~\cite{Ahlers:2005sn}
that certain low cross-over models which require very powerful sources
are already marginally excluded by experimental upper bounds on
diffuse neutrino fluxes. However, a general exclusion of such models
based on neutrino bounds alone has not yet been possible, given the
uncertainties inherent in the acceleration mechanism as well as the
possible cosmic evolution of the sources.

We update and improve this observation~\cite{Ahlers:2005sn} in several
respects. The bound on diffuse neutrino fluxes below $10^9$~GeV has
improved by a factor of 4 between the
AMANDA-B10~\cite{Ackermann:2005sb} and the
AMANDA-II~\cite{Ackermann:2007km,Gerhardt:2007zz} measurements and
will improve further by an order of magnitude after 1 year of
observation with IceCube which is currently under construction at the
south Pole~\cite{Achterberg:2006md}. In addition, the hybrid
measurements of the Pierre Auger Observatory have reduced the uncertainties in the absolute
normalization of the UHE CR flux~\cite{Abraham:2008ru} and these
measurements disagree with both the AGASA data~\cite{Takeda:2002at}
and the HiRes data~\cite{AbuZayyad:2002sf,Abbasi:2007sv}, which was used in the
earlier analysis~\cite{Ahlers:2005sn} (see also
Ref.~\cite{Ahlers:2007jc}). Furthermore, we will present a method to
derive {\em differential} upper limits on the proton fraction of UHE
CRs from experimental bounds on the diffuse neutrino flux.

The outline of this paper is as follows. We start in
Sect.~\ref{optthin} by deriving the luminosity relation between
neutrons and neutrinos in optically and magnetically thin sources. In
Sect.~\ref{boltzmann} we discuss the Boltzmann equations that govern
the propagation of UHE CRs and neutrinos. These results are used in
Sect.~\ref{protonfrac} to derive upper limits on the extra-galactic
proton contribution to UHE CRs from candidate sources such as
blazars. We summarise our findings in Sect.~\ref{conclusions} and
provide derivations of the results referred to in the main text in the
Appendices.

\section{Extra-galactic Neutrino Production}\label{optthin}

Cosmic rays originating in cosmic accelerators will typically be
accompanied by gamma-rays and neutrinos. This is a consequence of the
inelastic hadronic processes which are involved in their production
mechanism~\cite{Rachen:1998fd}. Candidate sources for the highest
energy CRs such as blazers or gamma ray bursts are expected to
accelerate charged particles by the 1st-order Fermi process or
`diffusive shock acceleration' in which they repeatedly scatter off a
propagating plasma shock front (see Ref.~\cite{Drury:1983zz}). (It is 
also possible that UHE CRs are accelerated by the 2nd-order Fermi process, 
{\it e.g.}~in the extended lobes of radio galaxies~\cite{Fraschetti:2008uc}.) 
For efficient acceleration up to the highest observed energies this
process has to be repeated many times hence it is essential that a
magnetic field confines the charged particles for sufficiently long in
the vicinity of the shock front.

Accelerated electrons and other light charged particles will lose
their energy in the magnetic field due to synchroton radiation. The
resulting photons provide a target for protons and heavier nuclei to
undergo meson photo-production and photo-disintegration,
respectively. A neutron produced in this process may escape out of
the magnetically confined source before it $\beta$-decays into a CR
proton. To maintain the efficiency of the acceleration mechanism, such
interactions need to be much less frequent than the acceleration
cycle. Charged and neutral pions decay further into high energy
neutrinos and gamma rays which are not confined by the magnetic field
and can be emitted from the source.

Here we focus on cosmic proton accelerators. Depending on the relative
ambient gas and photon densities, charged pion production may proceed
either through inelastic $pp$ scattering~\cite{Anchordoqui:2004eu}, or
photopion production predominantly through the resonant process
\mbox{$p\gamma \to \Delta^+\to n\pi^+$}~\cite{Waxman:1998yy}. The subsequent
decay chain \mbox{$\pi^+\to\mu^+\nu_\mu\to
e^+\nu_e\overline{\nu}_\mu\nu_\mu$} produces high energy neutrinos with
an initial flavor ratio $\nu_e\!:\!\nu_\mu\!:\!\nu_\tau$ of
$1\!:\!2\!:\!0$, which is expected to evolve through oscillations into
approximately $1\!:\!1\!:\!1$ after propagation over cosmological
distances~\cite{Learned:1994wg}.

The relative luminosity of neutrons and neutrinos depends on various
details of the source~\cite{Rachen:1998fd}. Since the interaction time
scale of protons and neutrons is similar, the relative luminosity of
neutrinos will tend to be {\em higher} in sources which are optically
thick to $(p/n)\gamma$ interactions. The life-time, geometry, bulk
motion, and magnetic field of the source limits the maximal energy of
the emitted protons~\cite{Hillas:1985is}. Also important are cooling
processes of the muons and pions in the source prior to decay, such as
adiabatic losses due to the expansion of the source, inverse Compton
scattering, Bethe--Heitler pair production, and synchrotron radiation
losses~\cite{Rachen:1998fd}.

Our working hypothesis in the following can be expressed in terms of
the time scales of $p\gamma$ and $pp$ interactions ($\tau_{pp}$ and
$\tau_{p\gamma}$), the decay lifetimes of neutron, pion and muon
($\tau_n$, $\tau_\pi$ and $\tau_\mu$), the cooling time scale of
charged particles ($\tau_{\rm cool}$), the characteristic cycle time
of confinement ($\tau_{\rm cycle}$), and the total confinement time
scale ($\tau_{\rm conf}$): ($i$) $\tau_{p\gamma} \ll \tau_{pp}$;
($ii$) $\tau_{\rm cycle}< \tau_n$; ($iii$) $\tau_{\rm
  cycle}\lesssim\tau_{p\gamma}$; ($iv$) $\tau_{p\gamma}\ll\tau_{\rm
  conf}$; ($v$) $\tau_{n/\pi/\mu} < \tau_{\rm cool}$.

Conditions ($i$) to ($iii$) ensure that photopion production of
neutrons and subsequent diffusion out of the magnetic confinement
region is the dominant process of cosmic proton production in an {\it
  optically thin} source. Conditions ($iii$) and ($iv$) permit
efficient cosmic ray acceleration with a smooth spectrum across a wide
range of energy. Finally, condition ($v$) ensures that cooling
processes of secondary neutrons, pions, and muons have negligible
influence on the energy distribution of the emitted neutrons and
neutrinos.

\begin{figure}[t]
\begin{center}
\includegraphics[width=\linewidth,clip=true,bb=161 497 500 678]{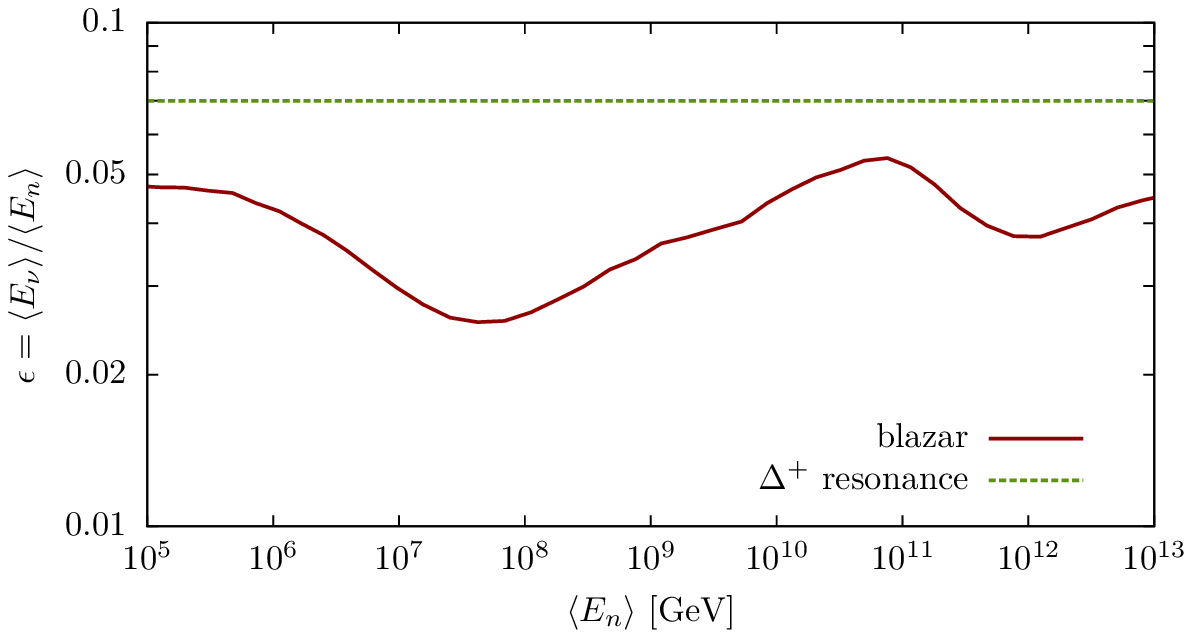}\\[0.2cm]
\includegraphics[width=\linewidth,clip=true,bb=154 496 500 680]{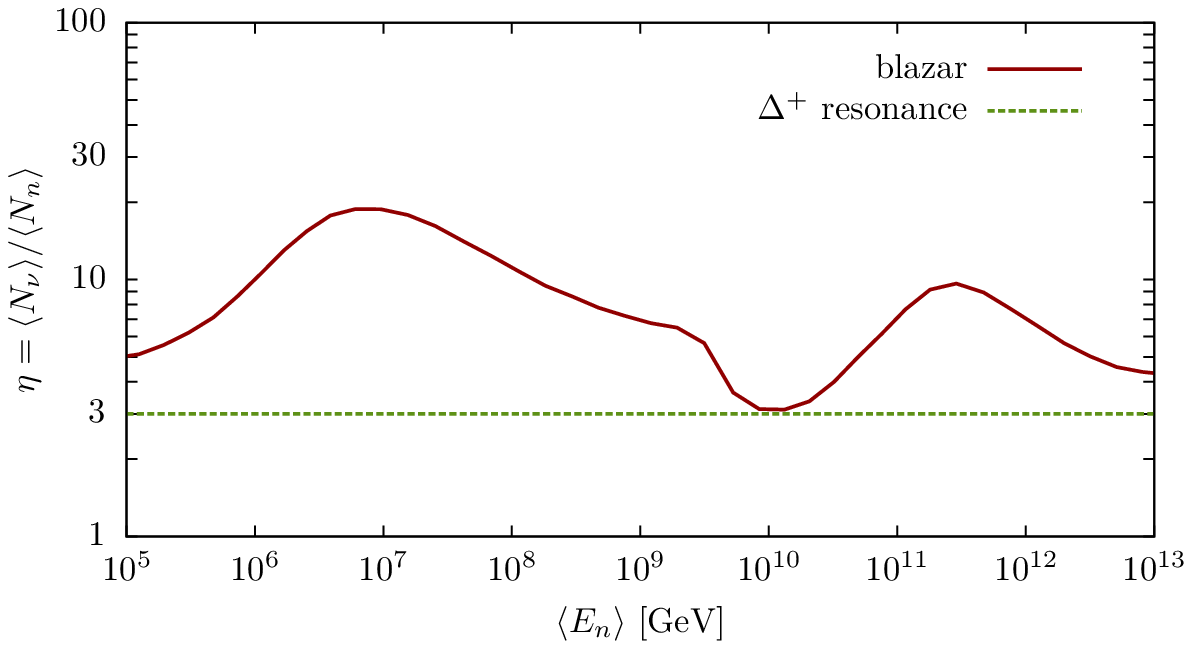}
\end{center}
\vspace{-0.3cm}
\caption[]{The values of $\epsilon = \langle E_\nu\rangle/\langle E_n
  \rangle$ and $\eta = \langle N_\nu\rangle/\langle N_n \rangle$ as
  calculated with the SOPHIA Monte Carlo code~\cite{Mucke:1999yb} for
  the blazar photon target spectrum shown in Fig.~1 of
  Ref.~\cite{Anchordoqui:2007tn}. For comparison we show the values at
  the $\Delta^+$ resonance used previously in
  Ref.\cite{Ahlers:2005sn}.}
\label{epseta}
\end{figure}

Within this working hypothesis the relative number and energy of the
neutrinos and neutrons depend only on the kinematics of photo-hadronic
interaction, which implies approximate equipartition of the decaying
pion's energy between the neutrinos and the electron, and the relative
radiation density in the source. On average, each proton-photon
interaction will produce $\eta$ neutrinos per neutron with relative
energy $\epsilon$ per neutrino, {\it i.e.}
\begin{equation}
  \eta = \frac{\langle N_\nu\rangle}{\langle N_n\rangle}\quad\text{and}\quad\epsilon = 
  \frac{\langle E_\nu\rangle}{\langle E_n\rangle}\,.
\end{equation}
In the following, we will consider a (hypothetical) source, where
photo-pion interaction proceed exclusively via the $\Delta^+$
resonance with fixed values $\eta=3$ and
$\epsilon=0.07$~\cite{Ahlers:2005sn}, and the blazar flaring state
model~\cite{Boettcher:1999ab} that has been previously discussed in
Ref.~\cite{Anchordoqui:2007tn}.

For a given photon target spectrum in the source the values of $\eta$
and $\epsilon$ can be obtained using the SOPHIA Monte Carlo
code~\cite{Mucke:1999yb} for photo-nucleon interactions and are shown
in Fig.~\ref{epseta} for the adopted blazar spectrum. For a fixed
proton energy we sample the scattering angle and photon energy in the
center-of-mass frame, weighted by the total $p\gamma$ cross section
and the photon distribution, {\it i.e.}~the integrand of
Eq.~(\ref{Gamma}). These parameters are then input to SOPHIA to
generate a $p\gamma$ interaction and the process is repeated $10^5$
times to derive the average values for the neutrino/neutron number and
energy.

The neutrino emissivity of flavor $i$ is then given by:
\begin{equation}
\frac{\Delta E_{\nu_i}}{N_{\nu_i}}\mathcal{L}_{\nu_i}(z,E_{\nu_i})
= \frac{\Delta E_n}{N_n}\mathcal{L}_n(z,E_n)\,.
\end{equation}
Assuming flavor universality as well as $\epsilon \simeq E_\nu/E_n
\simeq \Delta E_\nu/\Delta E_n$ and $\eta \simeq
N_{\mathrm{all}\,\nu}/N_n$ we arrive at the neutrino source luminosity
(per co-moving volume):
\begin{equation}\label{ratio}
\mathcal{L}_{\mathrm{all}\,\nu}(z,E_\nu) \simeq\frac{\eta}{\epsilon}\,
\mathcal{L}_n(z,E_\nu/\epsilon)\,.
\end{equation}
Note, that the relation~(\ref{ratio}) derived for optically thin
sources can be regarded as a {\em lower} limit on the neutrino
luminosity as long as energy-loss processes in the source are
negligible. We will illustrate this below for the various conditions
of our working hypothesis one by one.

Pion production may also proceed predominantly via inelastic $pp$
scattering if $\tau_{p\gamma} > \tau_{pp}$ ({\it cf.}~($i$)), in the
source. For the process $pp\to NN+{\rm pions}$, hadronic event
generators indicate an inelasticity $\sim 0.6$ where $N$ is a final
state nucleon~\cite{AlvarezMuniz:2002ne,Kelner:2006tc}. Assuming that 2/3 of the 
final state pions are charged we estimate that the average energy deposit 
into neutrinos is about $\eta\epsilon\sim1/4$ with $\eta\gg3$ due to the large multiplicity 
of secondary pions in inelastic $pp$ collisions. For a neutron luminosity $\mathcal{L}_n\propto E^{-\gamma}$ with \mbox{$\gamma\lesssim2$} (typical for relativistic plasma shocks) this will somewhat 
{\em increase} the neutrino flux relative to the neutrons 
({\it cf.}~Eq.~(\ref{ratio})).

If neutrons decay within the source, {\it i.e.}~$\tau_{\rm cycle} >
\tau_n$ ({\it cf.}~($ii$)), and the sources are optically thick, {\it
  i.e.}~$\tau_{\rm cycle}\gg\tau_{p\gamma}$ ({\it cf.}~($iii$)), then the
neutrino to neutron ratio would also be enhanced due to neutron
re-conversion into protons through $\beta$-decay and $n\gamma$
interactions, as well as simultaneous production of additional
neutrinos. The optical thickness of the source depends on the
intensity of the radiation field as well as the characteristic size
$R$ of the acceleration region. In particular, gamma-ray bursts,
possible sources of UHE CRs, are optically thick for proton energies
larger than $10^7$~GeV given standard
parameters~\cite{Piran:1999kx,Meszaros:2006rc} so we do not consider
direct neutrino emission from such sources in the following
(conservative) analysis.
 
However, if the target photon field gets too thin so that $\tau_{\rm
  cycle}\ll\tau_{p\gamma}$ ({\it cf.}~($iii$)), then the source would
rather accelerate protons to higher energy than emit neutrons through
photopion interactions. Whereas this does not change the ratio
(\ref{ratio}) and our analysis remains conservative, both neutron and
neutrino emission become inefficient. This is important for neutrino
energies below a few $10^8$~GeV in blazars~\cite{Anchordoqui:2007tn},
hence we will not use experimental neutrino bounds below $10^8$~GeV
for these candidate sources. However, one can envision cosmic ray
sources in which there is substantial leakage of protons from the
vicinity of the shock wave, which however remain trapped in the source
magnetic field until they interact with the photon background,
producing neutrons and neutrinos. Thus, in our general analysis of
optically thin sources we extend the energy range down to $10^6$~GeV.

Some care has to be taken if cooling processes of secondary pions and
muons in the source environment are in fact important
({\it cf.}~($v$))~\cite{Rachen:1998fd,Lipari:2007su}. Synchrotron
radiation, inverse Compton (IC) scattering, Bethe--Heitler (BH) pair
production, or adiabatic losses can reduce the energy of pions and
muons before decay and hence {\em decrease} the neutrino luminosity at
high energies. Inverse Compton cooling is the most relevant process
for photointeractions of pions and muons~\cite{Rachen:1998fd} and its
time scale can be related to the synchrotron loss time via the
relative co-moving energy densities of photons and the magnetic field,
$\tau^{\rm IC}_\star = (U_\gamma/U_B)\tau^{\rm sync}_\star = \tau^{\rm
  sync}_\star/\xi$. Here, we have introduced the equipartition
parameter $\xi$ which can exceed 1 in baryon-loaded flows due to the
low radiation efficiency of relativistic
protons~\cite{Rachen:1998fd}. However, Klein-Nishina corrections at
high pion and muon energies reduce the efficiency of IC cooling, which
might effectively lead to $\xi<1$~\cite{Lipari:2007su}. Hence, we will
estimate the time-scale of secondary particle cooling by $\tau^{\rm
  sync}_\star$.

Synchrotron losses of secondary pions and muons in the background
magnetic field are important if their lifetime $\gamma_\star\tau^{\rm
  dec}_\star$ is larger than the energy loss time $\tau^{\rm
  sync}_\star$, given by\footnote{We work throughout in natural
  Heaviside-Lorentz units with $\hbar=c=\epsilon_0=\mu_0=1$,
  $\alpha=e^2/(4\pi)\simeq1/137$ and $1~{\rm G} \simeq
  1.95\times10^{-2}{\rm eV}^2$.}
\begin{equation}
\tau^{\rm sync}_\star = \frac{9}{16\pi}\frac{m_\star^4}{\alpha^2E_\star B^2}\,.
\end{equation}
With muon and pion lifetimes of $\tau_\mu^{\rm dec} = 2.2\,\mu{\rm s}$
and \mbox{$\tau_\pi^{\rm dec} = 26\,{\rm ns}$}, respectively, this translates
into a critial energy in the observatory frame
\begin{gather}\nonumber
E^{\rm cr}_\star = \Gamma\frac{3}{4}\sqrt{\frac{m_\star^5}{\pi\alpha^2 B^2\tau_\star^{\rm dec}}}\\ =\left(\frac{\Gamma}{10^{1.5}}\right)\left(\frac{B}{10\,{\rm G}}\right)^{-1}\!\!\!\!\!\times\begin{cases}1.8\times10^{11}\,{\rm GeV}&\text{($\mu$)}\\3.4\times10^{12}\,{\rm GeV}&\text{($\pi$)}\end{cases},\label{secondloss}
\end{gather}
above which our relation between neutron and neutrino luminosities is unreliable. 

We use neutrino bounds from AMANDA
II~\cite{Ackermann:2007km,Gerhardt:2007zz} and
Auger~\cite{Abraham:2007rj}, as well as the projected sensitivity of
IceCube~\cite{Halzen:2006ic}, which extend up to energies of a few
$10^{10}$~GeV (in the observatory frame) and, after red-shifting,
correspond to neutrino emission in the source up to energies of about
$10^{11}$~GeV and pion/muon production up to a few $10^{11}$~GeV
(since the neutrinos carry only about 1/4 of the pion/muon
energy). Equation (\ref{secondloss}) shows that cooling processes in
blazar jet environments with typical boost factor
$\Gamma\simeq10^{1.5}$ and
$B\lesssim10$\,G~\cite{Biermann:1987ep,Anchordoqui:2001bs} may become
important for neutrinos from muon decay only at energies of about
$10^{10}$~GeV. However, it has been shown~\cite{Anchordoqui:2007tn}
that typically blazar sources become optically thick at these energies
which leads to an increase of the neutrino to proton ratio. In
contrast, GRBs typically have a much larger magnetic field of order
$B\simeq10^{12}$~G and a boost factor of
$\Gamma\simeq10^{2.5}$. Hence, cooling of secondaries in GRBs {\em
  cannot} be neglected, but as we have noted already they are not
optically thin sources at UHE and we do not therefore consider
(direct) neutrino emission from them in the following (conservative)
discussion.

The ratio of the observable flux of photo-hadronic neutrinos and CRs
today can be calculated from the relation (\ref{ratio}) using a set of
Boltzmann equations that take into account the interactions with the
cosmic photon background and the dilution and red-shift via cosmic
expansion. We will discuss these equations and their solutions in the
next section.

\begin{figure*}[t]
\begin{center}
\includegraphics[width=0.47\linewidth]{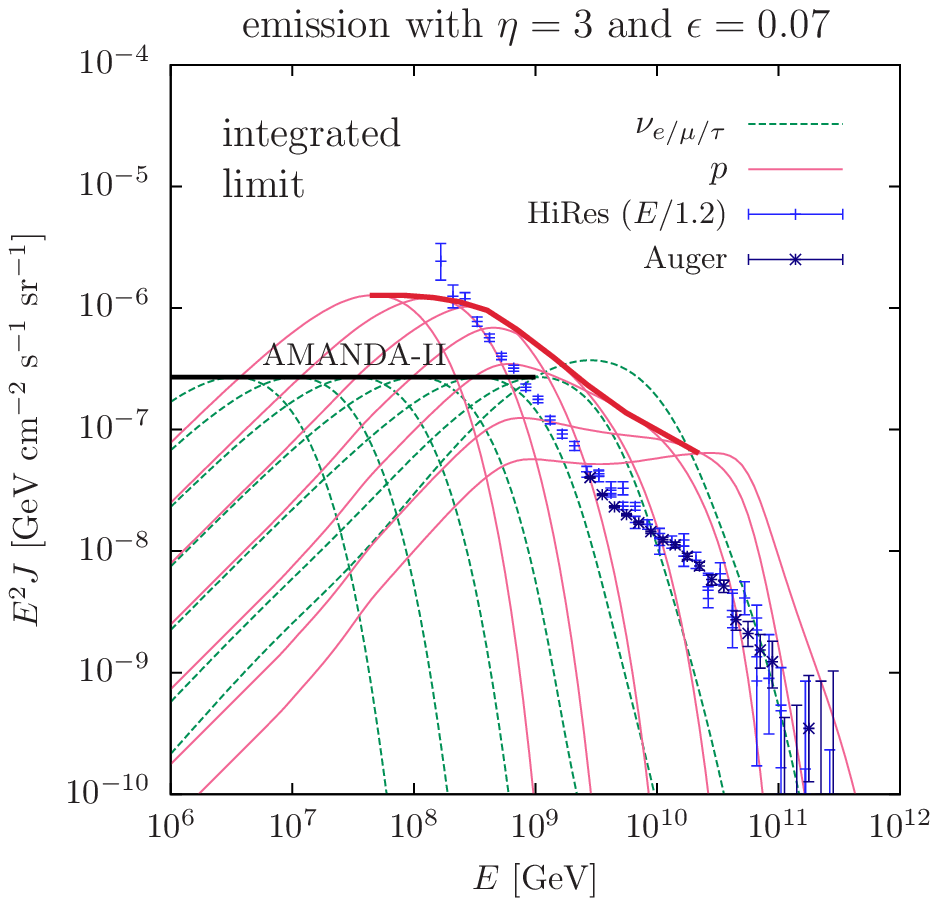}
\hspace{0.5cm}
\includegraphics[width=0.47\linewidth]{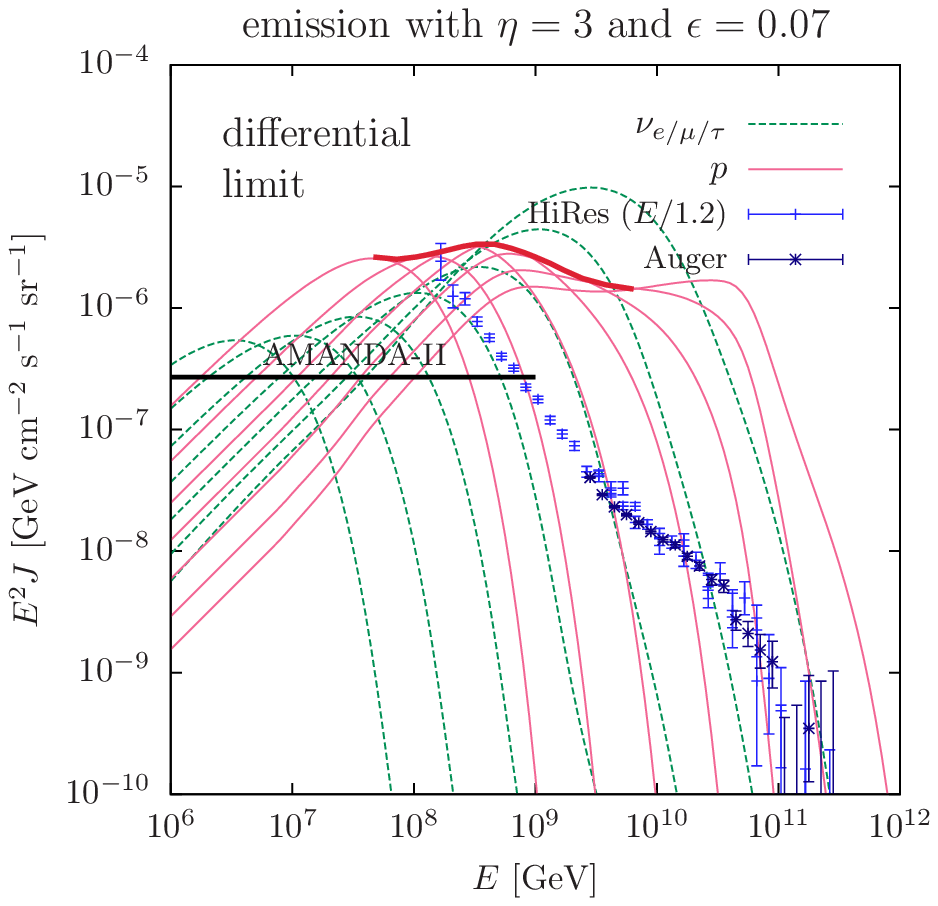}\\[0.3cm]
\includegraphics[width=0.47\linewidth]{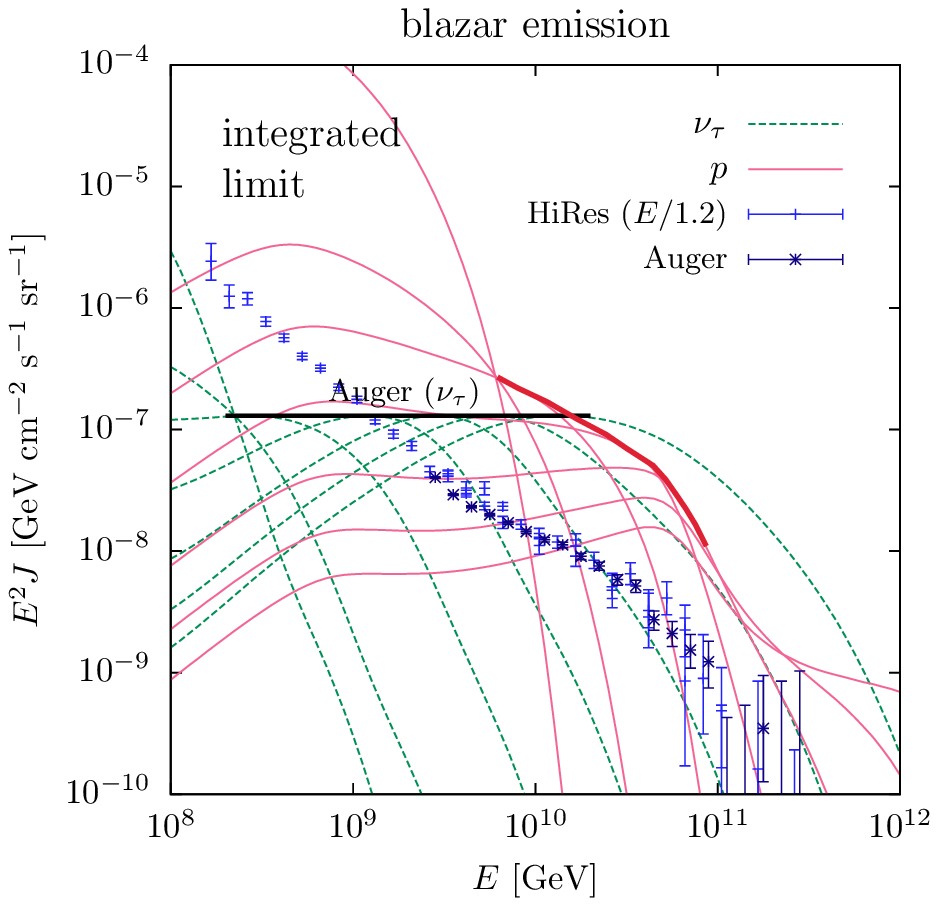}
\hspace{0.5cm}
\includegraphics[width=0.47\linewidth]{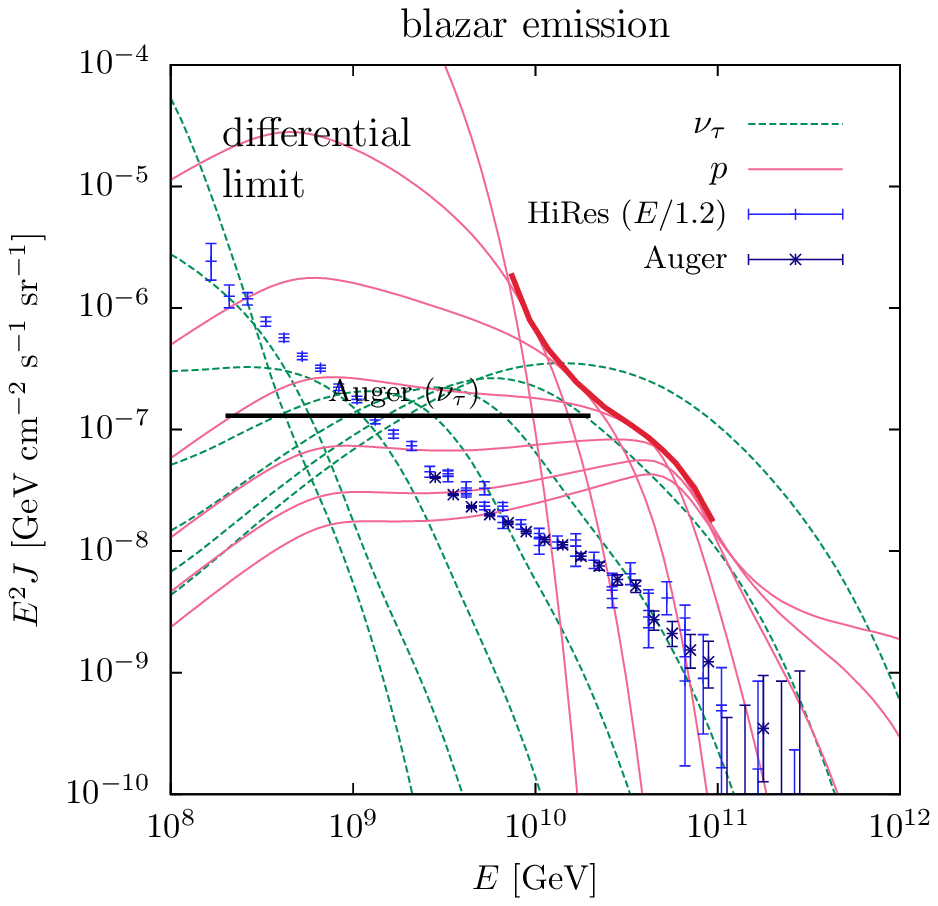}
\end{center}
\vspace{-0.3cm}
\caption[]{Integrated and differential upper limits on the proton
  contribution in UHE cosmic rays. The thin lines show a sample of proton 
  and neutrino test spectra (see Eqs.~(\ref{ratio}) and (\ref{test})) used 
  in the analysis. {\bf Upper panels}: Limits derived from the AMANDA-II bound on diffuse
  neutrinos~\cite{Ackermann:2007km}, $E^2J_{{\rm all}\, \nu} < 2.7
  \times10^{-7}$\,GeV\,cm${}^{-2}$\,s${}^{-1}$\,sr${}^{-1}$ (90\%
  C.L.) from $2\times10^5$~GeV to $10^9$~GeV, assuming that the
  sources are optically thin with $\eta=3$ and $\epsilon=0.07$
  ({\it cf.}~Ref.~\cite{Ahlers:2005sn}). {\bf Lower panels:} Limits derived
  from the Auger bound on UHE tau neutrinos~\cite{Abraham:2007rj},
  $E^2J_{\nu_{\tau}+\bar\nu_{\tau}} <
  1.3\times10^{-7}$\,GeV\,cm${}^{-2}$\,s${}^{-1}$\,sr${}^{-1}$ (90\%
  C.L.) from $2\times10^8$~GeV to $2\times10^{10}$~GeV, assuming that
  the sources are blazars. We adopt a neutrino flavor ratio
  $\nu_e\!:\!\nu_\mu\!:\!\nu_\tau$ of $1\!:\!1\!:\!1$ at Earth to
  exploit this bound for other flavours and to extract the
  differential limit we use the (relative) value of the exposure
  quoted in Ref.~\cite{BlanchBigas:2007tp}.}
\label{fraction}
\end{figure*}

\begin{figure*}[t]
\begin{center}
\includegraphics[width=0.47\linewidth]{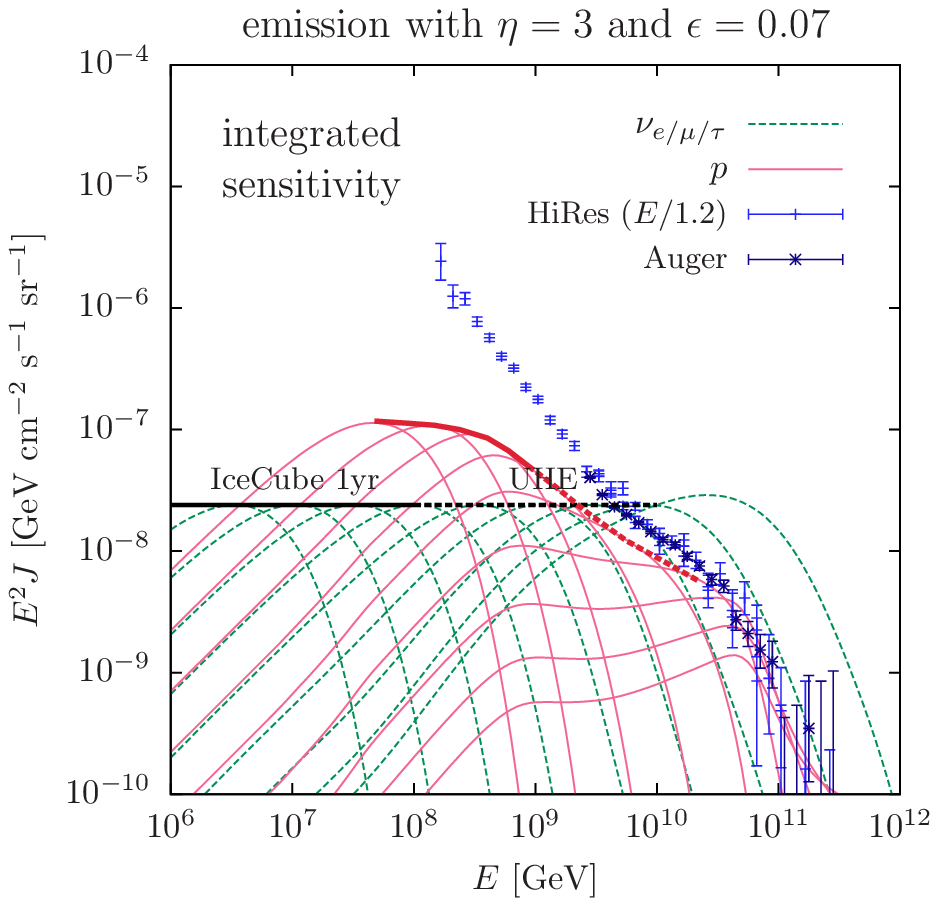}
\hspace{0.5cm}
\includegraphics[width=0.47\linewidth]{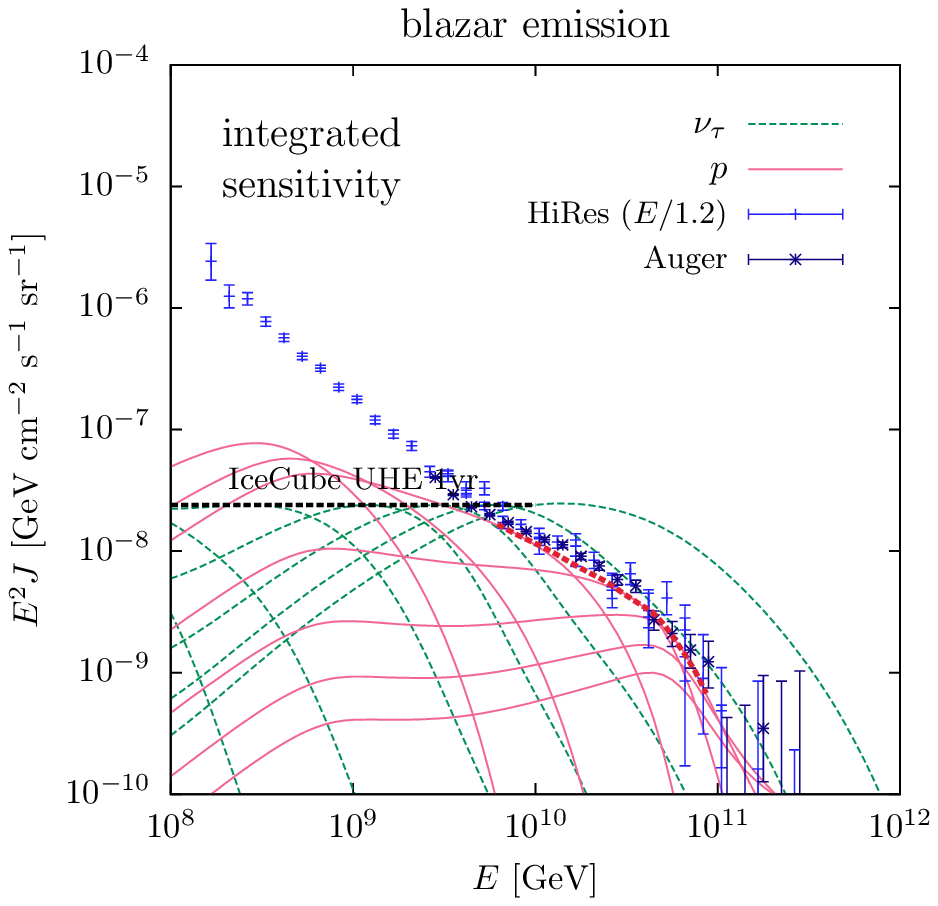}\\[0.3cm]
\includegraphics[width=0.47\linewidth]{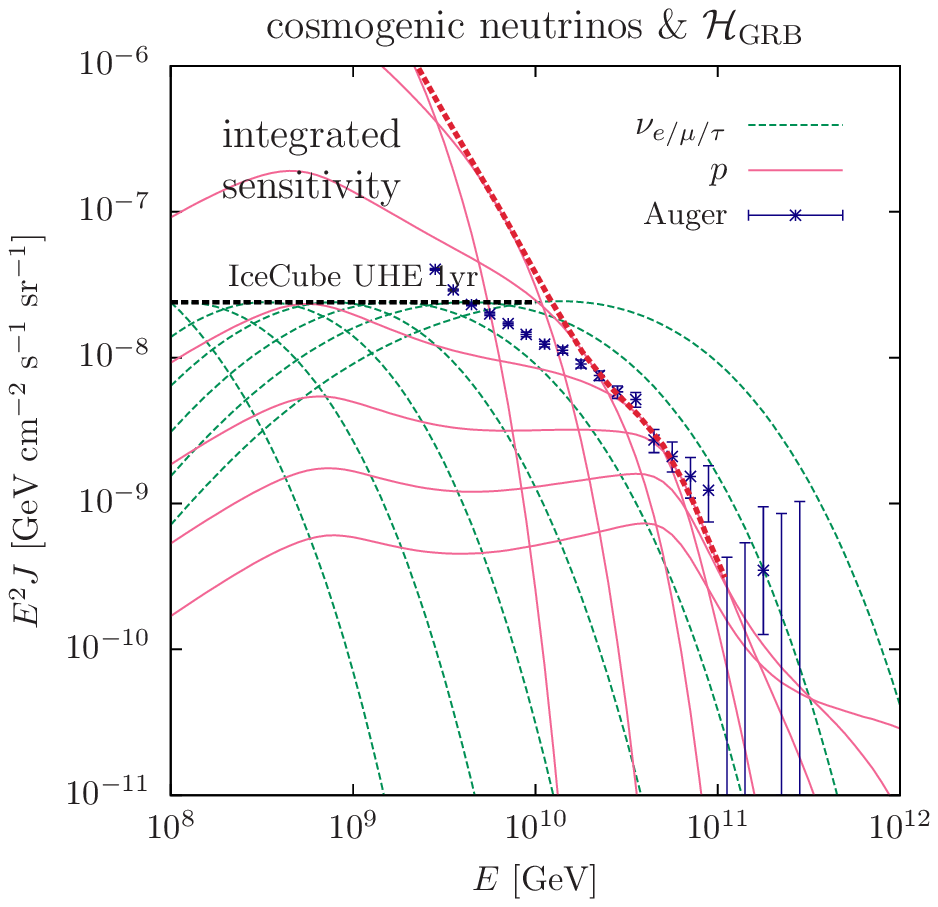}
\hspace{0.5cm}
\includegraphics[width=0.47\linewidth]{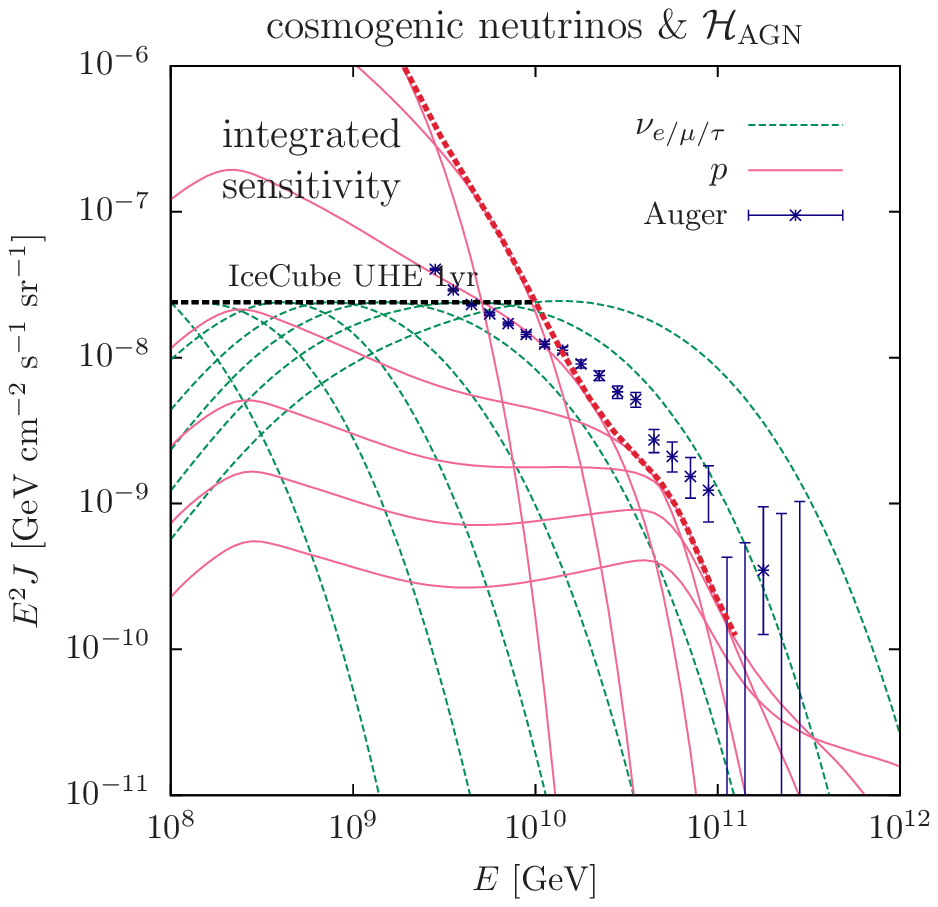}
\end{center}
\vspace{-0.3cm}
\caption[]{As in Fig.~\ref{fraction}, but now showing the estimated sensitivity 
  to the proton component in UHE CRs from the IceCube reach after one year of 
  observation. The IceCube sensitivity $E^2J_{{\rm all}\, \nu} \sim
  2.4\times10^{-8}$\,GeV\,cm${}^{-2}$\,s${}^{-1}$\,sr${}^{-1}$ on UHE
  neutrinos above $10^8$~GeV is estimated by an extrapolation to $10^{10}$~GeV
  motivated by the analysis of Ref.~\cite{Halzen:2006ic}. {\bf Upper
    panels:} Sensitivity to the proton component exploiting both
  cosmogenic neutrinos and neutrinos from optically thin sources. {\bf
    Lower panels:} Sensitivity exploiting cosmogenic neutrinos alone
  in models where the CR sources evolve strongly with redshift; we
  show the case of strong cosmological evolution of the proton sources
  according to $\mathcal{H}_{\rm GRB}$ (\ref{HGRB}, left panel) and
  $\mathcal{H}_{\rm AGN}$ (\ref{HAGN}, right panel).}
\label{sensitivity}
\end{figure*}

\section{Propagation of Cosmic Rays and Neutrinos}\label{boltzmann}

For a spatially homogeneous distribution of cosmic sources, emitting
UHE particles of type $i$, the co-moving number density $Y_i$ is
governed by a set of 1-dimensional Boltzmann equations of the form
\begin{multline}\label{diff0}
\dot Y_i = \partial_E(HEY_i) + \partial_E(b_iY_i)\\-\Gamma_{i}\,Y_i+\sum_j\int\mathrm{d} E_j\,\gamma_{ji}Y_j+\mathcal{L}_i\,,
\end{multline}
together with the Friedman-Lemaitre equations describing the cosmic
expansion rate $H(z)$ as a function of the redshift
$z$. Non-relativistic and non-interacting matter obeys the equation
$\dot Y=0$. The first and second term in the r.h.s.~of
Eq.~(\ref{diff0}) describe continuous energy losses (CEL) due to
red-shift and $e^{+} - e^{-}$ pair production on the cosmic photon
backgrounds, respectively. The third and fourth terms describe more
general interactions involving particle losses ($i \to$ anything) with
interaction rate $\Gamma_i$, and particle generation of the form $j\to
i$.\footnote{We will illustrate in Appendix~\ref{appendixBI} that CEL
  processes are an approximate formulation of the process $i\to i$.}
The last term on the r.h.s.~of Eq.~(\ref{diff0}), $\mathcal{L}_i$,
corresponds to the luminosity density per co-moving volume of sources
emitting CRs of type $i$. In Appendices \ref{appendixBI},
\ref{appendixND} and \ref{appendixIR} we provide more details
regarding the quantities appearing in Eq.~(\ref{diff0}).

Note, that the Boltzmann equations~(\ref{diff0}) do not take into
account the deflection of charged CRs during their propagation through
inter-galactic and galactic magnetic fields. In fact, if synchrotron
radiation during propagation is negligible and the source distribution
is homogenous, Eq.~(\ref{diff0}) provides a good approximation of the
spectral evolution even for CRs having small rigidity which suffer
large deflections~\cite{Aloisio:2004jda}. However, magnetic
inhomogeneities on small scales will suppress the spectrum of CRs with
Larmor radius $\ell_\mathrm{L}<\ell_\mathrm{d}$ where $\ell_{\rm d}$
is the characteristic distance between sources. It has been shown that
for typical inter-galactic magnetic fields of strength $\sim 1$~nG and
coherence length of $\sim 1$ Mpc, the diffusive propagation of CR
protons will start to affect the spectrum below about $10^9$~GeV if
$\ell_\mathrm{d} \sim 50$~Mpc~\cite{Aloisio:2004fz}. Depending on the
diffusion regime, this can suppress the proton flux at $10^8$~GeV by a
factor of 3 to 100. We will return to this point in the discussion of
our results.

We adopt the usual concordance cosmology~\cite{Amsler:2008zzb} of a
flat universe dominated by a cosmological constant with
$\Omega_{\Lambda} \sim 0.7$, the rest being cold dark matter with
$\Omega_\mathrm{m} \sim 0.3$. The Hubble parameter is given by \mbox{$H^2
(z) = H^2_0\,(\Omega_{\mathrm{m} } (1 + z)^3 + \Omega_{\Lambda})$},
normalised to its value today of 70 km\,s$^{-1}$\,Mpc$^{-1}$. The
time-dependence of the red-shift can be expressed via $\mathrm{d}z =
-\mathrm{d} t\,(1+z)H$.

The cosmological evolution of the source density per co-moving volume
is parameterized as:
\begin{equation}\label{factorize}
\mathcal{L}_i(z,E) = \mathcal{H}(z)\mathcal{L}_i(0,E)\,,
\end{equation}
where the source luminosity per co-moving volume is assumed to follow
the star formation rate (SFR).  (Note that the dilution of the source
density due to the Hubble expansion is taken care of since
$\mathcal{L}$ is the {\em comoving} density, hence for no evolution we
would simply have $\mathcal{H} = 1$.)  Following the recent
compilation~\cite{Hopkins:2006bw,Yuksel:2008cu} we adopt
\begin{equation}\label{HSFR}
\mathcal{H}_{\rm SFR}(z) = \begin{cases}(1+z)^{3.4}&z<1\,,\\N_1\,(1+z)^{-0.3}&1<z<4\,,\\N_1\,N_4\,(1+z)^{-3.5}&z>4\,,\end{cases}
\end{equation}
with appropriate normalization factors, $N_1 = 2^{3.7}$ and $N_4 =
5^{3.2}$ (see the right panel of Fig.~\ref{scales}).

Some candidate sources of UHE CRs may have a stronger evolution than
the star formation rate (\ref{HSFR}); this is particularly important
for cosmogenic
neutrinos~\cite{Yuksel:2006qb,Takami:2007pp,Stanev:2008un}. For
gamma-ray bursts we adopt~\cite{Yuksel:2006qb}
\begin{equation}\label{HGRB}
\mathcal{H}_{\rm GRB}(z) = (1+z)^{1.4}\, \mathcal{H}_{\rm SFR}(z)\,.
\end{equation}
Active galactic nuclei may have a similarly strong
evolution. Following Refs.\cite{Hasinger:2005sb,Stanev:2008un} we take
\begin{equation}\label{HAGN}
\mathcal{H}_{\rm AGN}(z) = \begin{cases}(1+z)^{5.0}&z<1.7\,,\\N_{1.7}&1.7<z<2.7\,,\\N_{1.7}\,N_{2.7}^{(2.7-z)}&z>2.7\,,\end{cases}
\end{equation}
with $N_{1.7}=2.7^5$ and $N_{2.7}=10^{0.43}$. We will discuss the
impact of these strong evolution scenarios on cosmogenic neutrinos and
the proton fraction at very high energies.

In Appendix \ref{appendixBE} we derive a general solution to the
differential equation~(\ref{diff0}).  The neutrino flux from the
sources is obtained simply by integrating Eqs.~(\ref{diff1}) and
(\ref{diff2}):
\begin{equation}\label{nuflux}
J_\nu(E) = \frac{1}{4\pi}\int_0^\infty\D z\frac{1}{H(z)}\mathcal{L}_\nu(z,(1+z)E)\,.
\end{equation}
To obtain the cosmogenic neutrino flux produced during propagation
through cosmic radiation backgrounds we need to solve the full
Boltzmann equations numerically.

\section{Bounds on the Extra-Galactic Proton Fraction}\label{protonfrac}

The luminosity relation~(\ref{ratio}) between neutrinos and CRs in
optically thin sources was first used~\cite{Waxman:1998yy} to derive
an upper limit on the diffuse neutrino background from UHE CR
observations. This argument was modified and extended in further
studies which examined if the upper limit on the neutrino flux could
in principle be {\em exceeded}~\cite{Mannheim:1998wp,Ahlers:2005sn}.
With the advent of km${}^3$-scale neutrino telescopes like IceCube
these limits will soon be tested~\cite{Ahlers:2005sn}.

As we have outlined in Section \ref{optthin}, the relative neutrino
luminosity~(\ref{ratio}) of optically thin sources with photo-hadronic
neutrino production (conditions ($i$) to ($iii$)) can be regarded as a
{\it lower} limit on the neutrino flux, if cooling processes of
secondary pions and muons are negligible ($iv$) and proton diffusion
out of the confinement region is suppressed ($v$). We translate this
conservative expectation into an {\em upper} limit on the
extra-galactic proton fraction in UHE CRs, exploiting experimental
upper bounds on the diffuse high-energy neutrino flux.

For this procedure it is convenient to introduce test functions of the
neutron source luminosity of the form
\begin{equation}\label{test}
\mathcal{L}^{\rm test}_n(0,E) = \mathcal{L}_0\left(\frac{E}{E_\mathrm{max}}\right)^{-1}\!\!\!\exp\left(-\frac{E}{E_\mathrm{max}}\right)\,,
\end{equation}
with an exponential energy cut-off $E_\mathrm{max}$ that we vary
between $10^{8}$~GeV and $10^{12}$~GeV with a logarithmic step-size of
$\log_{10}E = 0.25$. Unless otherwise stated, we use
Eqs.~(\ref{factorize}) and (\ref{HSFR}) for the source evolution. Each
neutron test luminosity~(\ref{test}) is related to a neutrino
luminosity by the ratio~(\ref{ratio}). After propagation using
Eqs.~(\ref{diff0}) we normalize the accumulated contribution of
extra-galactic and cosmogenic neutrinos to the limit on the diffuse
neutrino flux from AMANDA II~\cite{Ackermann:2007km} and to the limit
on UHE tau neutrinos from Auger~\cite{Abraham:2007rj}. This can be done
in two possible ways.
 
For an {\em integrated} upper limit (or sensitivity) we maximise the
individual neutrino flux normalization to saturate the integrated
experimental neutrino flux bounds. This approach is shown in the left
panels of Fig.~\ref{fraction}. The envelope of the corresponding
proton flux gives the integrated upper limit on the proton
fraction. However, the application of this limit requires that the
flux of cosmic neutrinos is close to an $E^{-2}$ spectrum in the
quoted interval $[E_-,E_+]$. Note, that for CR protons above a few
times $10^8$~GeV this upper limit becomes trivial, {\it i.e.}~it lies
{\em above} the observed CR flux.

A {\em differential} upper limit can be obtained using the effective
area $A_\mathrm{eff}$ of AMANDA-II, provided in
Ref.~\cite{Gerhardt:2007zz}, and the acceptance ($\propto
A_\mathrm{eff}$) of Auger quoted in Ref.~\cite{BlanchBigas:2007tp}. The
normalisation of the neutrino test fluxes $J^\mathrm{test}_\nu$ is
fixed by the integrated limit $J^\mathrm{int}_\nu$ and the equation
\begin{equation}
\int\limits_{E_-}^{E_+} \D E\,A_{\rm eff} J^{\rm int}_\nu = \int\limits_{E_-}^{E_+} \D E\,A_{\rm eff} J^{\rm test}_\nu\,,
\end{equation}
with the appropriate integration limits $E_\pm$ according to the
experimental integrated neutrino flux bound. This second approach is
shown in the right panels of Fig.~\ref{fraction}. Again, the envelope
of the corresponding proton fluxes provides the differential upper
limit on the proton fraction. Note, that this differential bound is
more general and does not make the usual assumption that the neutrino
energy spectrum is $\propto E^{-2}$, but is in general weaker than the
integrated limit by up to a factor of 10.

One year of observation with the fully deployed IceCube detector
should increase the sensitivity to diffuse neutrino fluxes by about an
order of magnitude below $10^{8}$~GeV. It is conceivable that this
sensitivity can be extended to UHE neutrino energies of about
$10^{10}$~GeV~\cite{Halzen:2006ic}. The corresponding (integrated)
sensitivities on the proton fraction in UHE CRs are shown in the upper
panels of Fig.~\ref{sensitivity}. We see that 1 year of observation
is sufficient to place stringent bounds on the extra-galactic proton
fraction in CRs below $10^{10}$~{\rm GeV} and thus provide a
definitive test of the `dip-transition'
model~\cite{Berezinsky:2002nc}.

As noted earlier, the interplay of a random walk of cosmic protons in
magnetic fields together with an inhomogenous distribution of their
sources may result in suppression of the observed proton flux below
about $10^9$~GeV~\cite{Aloisio:2004fz}. Hence, we expect that our
upper limit on the proton flux in this energy region is rather
conservative. In addition, a source evolution much stronger than the
star formation rate (\ref{HSFR}) would predict a much higher flux of
neutrinos -- cosmogenic as well as from the sources -- whereas the
spectrum of protons is fixed by the contribution of local sources
within the `GZK horizon'.

The flux of cosmogenic neutrinos, which dominates the total neutrino
flux above $10^{10}$~GeV, can also in principle be exploited to obtain
bounds on the proton fraction independent of the nature of the
sources~\cite{Takami:2007pp}. However, since the CR flux is steeply
falling as $\sim E^{-3}$, to constrain the contribution of trans-GZK
protons would require even more sensitive experiments than IceCube
and/or a significantly stronger source evolution with redshift than we
believe is plausible. In the lower panels of Fig.~\ref{sensitivity} we
estimate the sensitivity of IceCube to constrain the proton fraction
using only cosmogenic neutrinos for two strong evolution scenarios
(see Eqs.~(\ref{HGRB}) and (\ref{HAGN})). Again, the IceCube
sensitivity shown to UHE neutrinos up to $10^{10}$~GeV is extrapolated
from lower energies, motivated by the analysis of
Ref.~\cite{Halzen:2006ic}. We observe that 1 year of observation at
IceCube would be sufficient to constrain trans-GZK protons in UHE CRs
only if the source evolution is sufficiently strong: $\propto(1+z)^5$
or steeper.

\begin{figure}[t]
\begin{center}
\includegraphics[width=\linewidth]{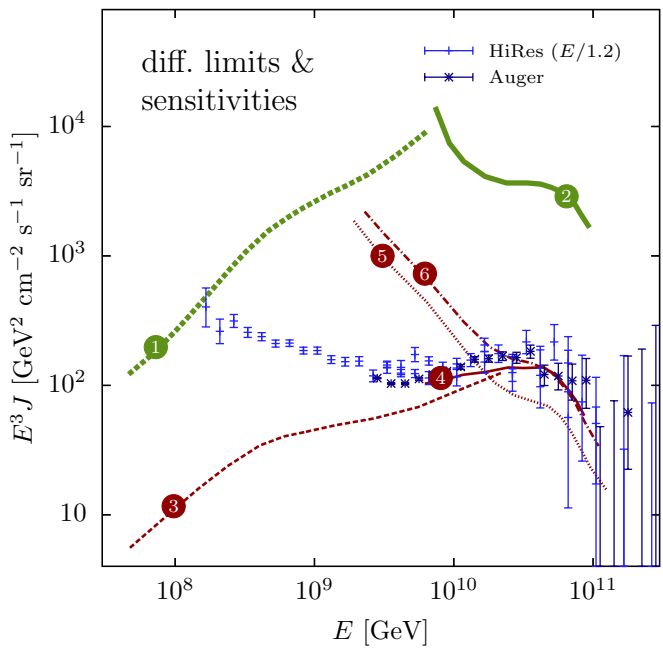}
\end{center}
\vspace{-0.5cm}
\caption[]{Summary of differential upper limits (thick lines 1\&2) and 
  integrated sensitivities (thin lines 3-6) on the extra-galactic proton 
  contribution compared to CR data from HiRes-I/II (recalibrated) and the 
  Pierre Auger Observatory (see Figs.~\ref{fraction} and \ref{sensitivity} 
  for details). The limits and sensitivities are derived from the $\Delta^+$ 
  approximation (1\&3), from blazar emission (2\&4) and from cosmogenic 
  neutrinos with a strong source evolution $\mathcal{H}_{\rm AGN}$ (5) and 
  $\mathcal{H}_{\rm GRB}$ (6), respectively.}
\label{bounds}
\end{figure}

\section{Conclusions}\label{conclusions}

Exactly where the transition occurs between the galactic and
extra-galactic components of UHE cosmic rays is presently an open
question and identifying this energy would provide important clues to
unravelling their origin. The acceleration of CRs in extra-galactic
sources would be accompanied by the emission of high energy neutrinos,
thus an upper limit on the proton flux can in principle be inferred
from experimental bounds on the extra-galactic UHE neutrino flux. We
have demonstrated this, focussing on blazar jets which we have argued
are optically thin sources, so photo-hadron interactions generate a
certain minimum flux of UHE cosmic neutrinos as long as proton
diffusion and secondary particle cooling are negligible. We have shown
that this argument is conservative in that a {\em higher} neutrino
flux would be expected if any of our assumptions are relaxed, {\it
  e.g.} if the sources of UHE CRs are optically thick.

Our main results are summarized in Fig.~\ref{bounds}. The AMANDA-II
bound on diffuse neutrinos already constrains the extra-galactic proton contribution
in CRs at energies below a few times $10^8$~GeV and just 1 year of
observation with IceCube will provide the necessary sensitivity up to
a few times $10^{10}$~GeV. If the number density of extra-galactic CR
sources evolves strongly with redshift then the detection of
cosmogenic neutrinos alone may enable bounds to be placed on the
proton flux at trans-GZK energies.

\section*{Acknowledgements}

MA acknowledges support by STFC UK (PP/D00036X/1). The research of LAA
has been partially supported by the US National Science Foundation
(Grant PHY-0757598) and UWM's RGI.  SS acknowledges a PPARC Senior
Fellowship (PPA/C506205/1) and support from the EU Marie Curie Network
``UniverseNet'' (HPRN-CT-2006-035863).

\begin{appendix}

\begin{figure*}[t]
\begin{center}
\includegraphics[width=0.47\linewidth,clip=true,bb=204 465 470 706]{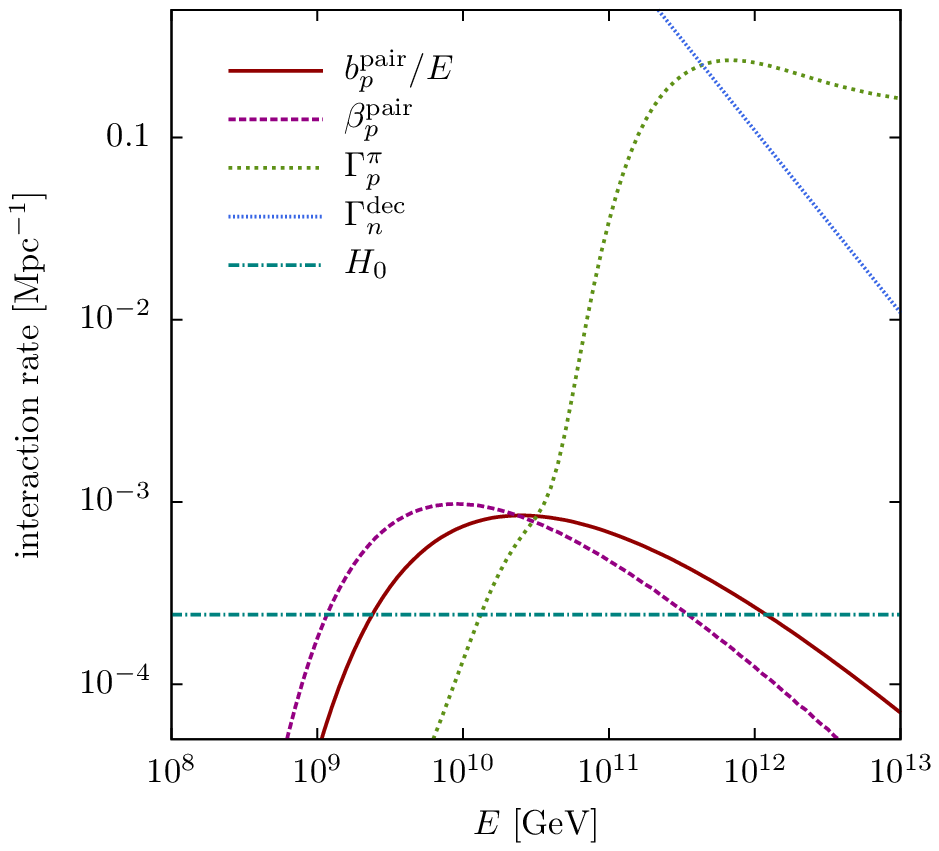}
\hspace{0.5cm}
\includegraphics[width=0.47\linewidth,clip=true,bb=185 465 451 710]{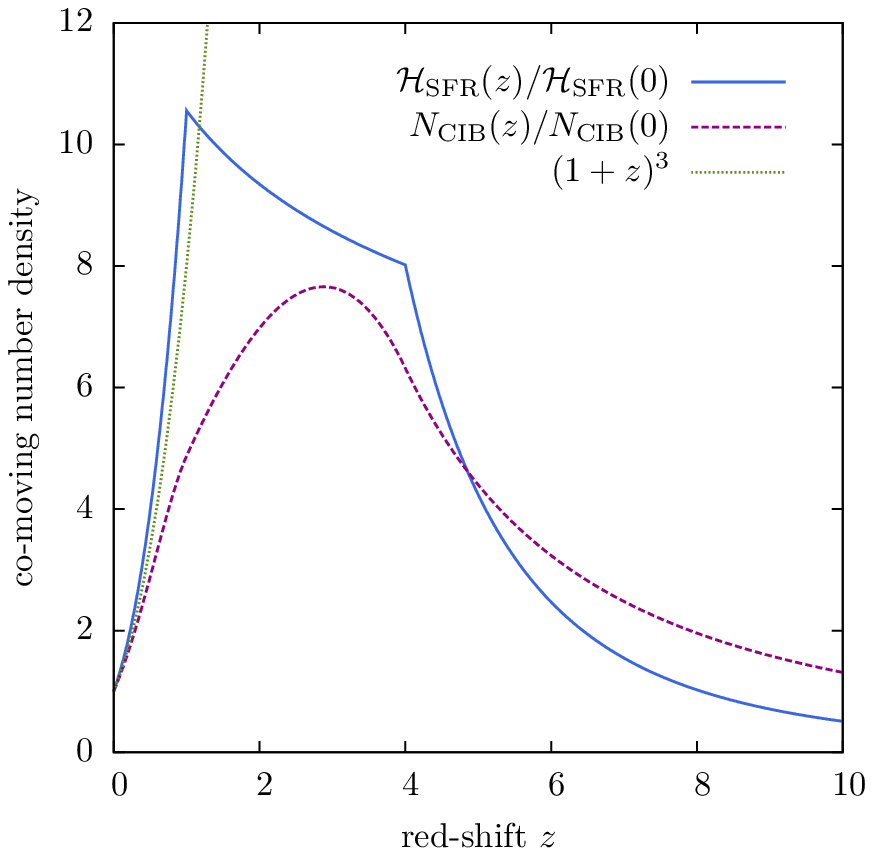}
\end{center}
\vspace{-0.3cm}
\caption[]{{\bf Left Panel:} The interaction and decay rates appearing
  in the Boltzmann equations for the CMB and
  CIB~\cite{Franceschini:2008tp} at $z=0$. {\bf Right Panel:} Star
  formation rate (Eq.~(\ref{HSFR}) from Ref.~\cite{Yuksel:2008cu}) and
  our approximation of the CIB number density scaling with redshift
  (\ref{HIR}). For comparison, we also show the scaling behaviour of
  the CMB number density $\propto(1+z)^3$.}
\label{scales}
\end{figure*}

\section{Photon Background Interactions}\label{appendixBI}

The propagation of UHE cosmic nuclei is affected by photo-hadronic
interactions on cosmic photon backgrounds. For UHE protons the
dominant interactions that determine the spectrum occur on the CMB,
but the cosmic infra-red background (CIB) is also important for the
generation of cosmogenic neutrinos. For the spectrum of the latter we
use the recent compilation of Ref.~\cite{Franceschini:2008tp}.

The angular-averaged (differential) interaction rate $\Gamma_i$
($\gamma_{ij}$) appearing on the r.h.s.~of Eq.~(\ref{diff0}) is
defined as
\begin{gather}\label{Gamma}
\Gamma_{i}(z,E_i) = \frac{1}{2}\int\limits_{-1}^1\mathrm{d}{\rm c}_\theta\int\mathrm{d}\epsilon\,(1-\beta {\rm c}_\theta) n_\gamma(z,\epsilon)\sigma^\mathrm{tot}_{i\gamma}\,,\\
\gamma_{ij}(z,E_i,E_j) = \Gamma_i(z,E_i)\,\frac{\D N_{ij}}{\D E_j}(E_i,E_j)\,,
\label{gamma}
\end{gather}
where $n_\gamma(z,\epsilon)$ is the energy distribution of background
photons at redshift $z$ and $\D N_{ij}/\D E_j$ is the angular-averaged
distribution of particles $j$ after interaction. For photo-hadronic
interactions this distribution can be determined using the Monte Carlo
package SOPHIA~\cite{Mucke:1999yb}. The factor $1-\beta{\rm c}_\theta$
takes into account the relative motion of photons and the nucleus,
{\it i.e.}~the Doppler shift of the photon density.

We assume that the photon background has the adiabatic scaling behaviour:
\begin{equation}\label{CMB}
n_\gamma(z,\epsilon) = (1+z)^3\,n_\gamma(0,\epsilon/(1+z))\,.
\end{equation}
This is exact for the CMB (following from \mbox{$\dot
Y_\gamma=\partial_E(HEY_\gamma)$} and $Y_\gamma \propto a^3n_\gamma$),
but not so for the CIB. However, the dominant opacity for proton
propagation is provided by the CMB.  The scaling behaviour
Eq.~(\ref{CMB}) translates into the following scaling of the
quantities $\Gamma_i$ and $\gamma_{ij}$,
\begin{align}\label{scaling1}
\Gamma_i(z,E_i) &= (1+z)^3\,\Gamma_i(0,(1+z)E_i)\,,\\\label{scaling2}
\gamma_{ij}(z,E_i,E_j) &= (1+z)^4\,\gamma_{ij}(0,(1+z)E_i,(1+z)E_j)\,.
\end{align}
For the adopted scaling of the CIB see Appendix~\ref{appendixIR}.

If all interactions can be described as a CEL process the differential
equation (\ref{diff2}) is considerably simplified and can be solved in
a closed form as we will show later. In general, any transition $i\to
i$ which can be approximated as
$\gamma_{ii}(E,E')\approx\delta(E-E'-\Delta E)\Gamma_i(E)$ with
$\Delta E/E \ll 1$ can be replaced in the Boltzmann
equations~(\ref{diff0}) as
\begin{multline}
-\Gamma(E)Y_i(E)+\int\mathrm{d} E'\, \gamma_{ii}(E',E)\,Y_i(E')\\\to\partial_E(b_i Y_i)\,, 
\end{multline}
with $\displaystyle b_i \equiv \Delta E\,\Gamma_i \approx -\dot E$.
The production of electron-positron pairs in the photon background
with a small energy loss is usually approximated as a CEL
process. Here we follow the standard approach of
Ref.~\cite{Blumenthal:1970nn} to calculate the proton energy losses in
the photon background. Again, the computation of the quantity $b$ at
various red-shift is significantly simplified if we assume an
adiabatically scaling background photon density as for the CMB. The
scaling behaviour of $b$ and its derivative $\beta=\partial_Eb$ is
then
\begin{align}\label{scaling3}
b_i(z,E_i) &= (1+z)^2\,b_i(0,(1+z)E_i)\,,\\\label{scaling4}
\beta_i(z,E_i) &= (1+z)^3\,\beta_i(0,(1+z)E_i)\,.
\end{align}
As before, for the scaling of the infra-red background see
Appendix~\ref{appendixIR}. In the left panel of Fig.~(\ref{scales}) we
show the quantities $b^\mathrm{pair}_p/E$, $\beta^\mathrm{pair}$,
$\Gamma_p$ and $H_0$ for comparison.

\section{Neutron Decay}\label{appendixND}

For neutrons produced by proton interactions on the photon background,
the Boltzmann equations have to include a decay term. Since the proton
and neutron are much heavier than the electron and electron
anti-neutrino we can safely assume that the proton is at rest in the
center of mass frame and $E^*_{\bar\nu_e}+E^*_e \approx m_n-m_p =
\Delta m$. The energy distribution of electron and electron
anti-neutrino in the nucleon's rest frame can then be approximated as
\begin{align}
  \frac{\mathrm{d}N_e}{\mathrm{d}E^*} &\propto\sqrt{{E^*}^2-m_e^2}E^*(\Delta m-E^*)^2\,,\\
  \frac{\mathrm{d}N_{\bar\nu}}{\mathrm{d}E^*} &\propto\sqrt{(\Delta
    m-E^*)^2-m_e^2}{E^*}^2(\Delta m-E^*)\,,
\end{align}
according to the phase space density of electron and anti-neutrino
with $\D N \sim \D^3p_e\D^3p_{\bar\nu_e}$ and the constraint \mbox{$\Delta m
= E_e+E_{\bar\nu_e}$}. The angular-averaged distribution in the lab
frame is
\begin{equation}
  \frac{\mathrm{d}N}{\mathrm{d}E} = \frac{1}{2}\int\limits_{-1}^1\,\mathrm{d}{\rm c}_\theta\, \frac{\mathrm{d}E^*}{\mathrm{d}E}\frac{\mathrm{d}N}{\mathrm{d}E^*}\,,
\end{equation}
with $E=\gamma_nE_*-\gamma_n\beta_n{\rm c}_\theta p_*$. The energy of
the emerging proton is approximately $\gamma_n m_p = m_pE_n/m_n\approx
E_n$.

\begin{figure*}[t!]
\begin{center}
\includegraphics[width=0.47\linewidth,clip=true,bb=203 465 468 707]{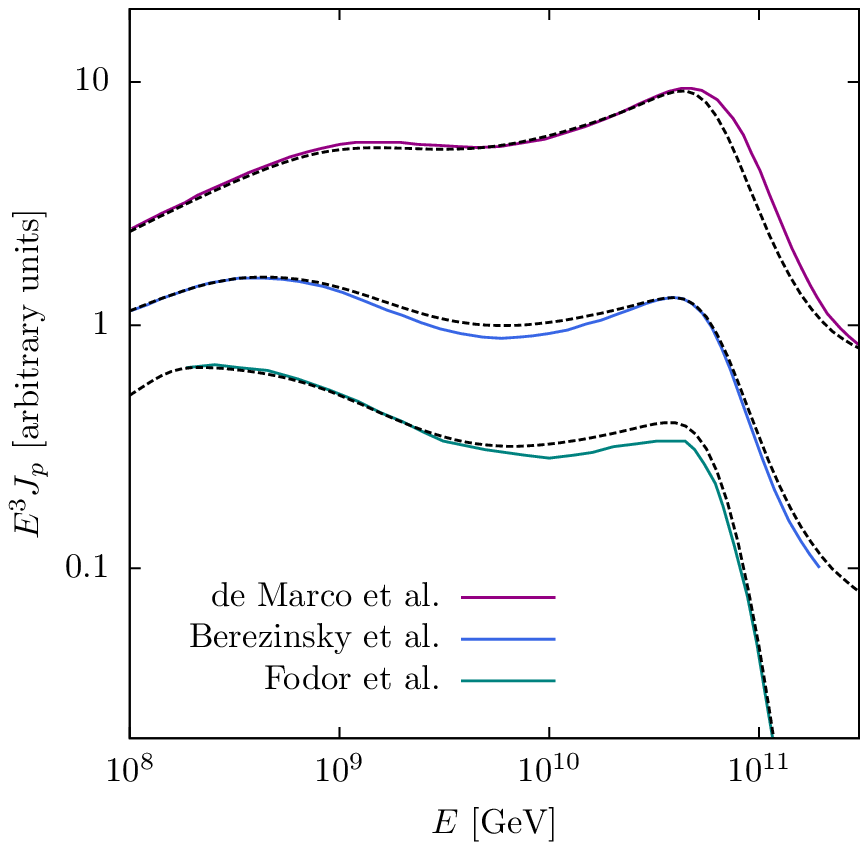}
\hspace{0.5cm}
\includegraphics[width=0.47\linewidth,clip=true,bb=207 465 472 711]{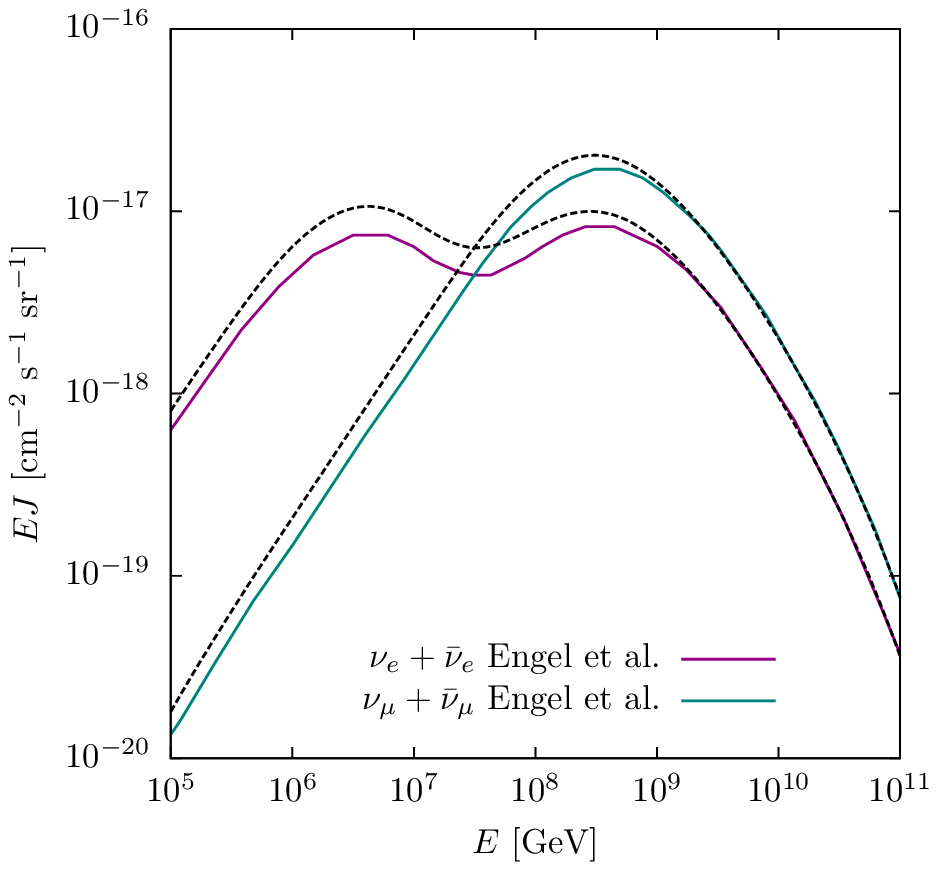}
\end{center}
\vspace{-0.3cm}
\caption[]{Comparison of our results (dashed lines) for the
  extra-galactic proton and cosmogenic neutrino flux with previous
  calculations. {\bf Left Panel:} Proton spectra from (solid lines;
  top to bottom) Ref.~\cite{DeMarco:2005ia} (their Fig.~6 with $n=0$,
  $\gamma=2.55$, $E_\mathrm{max}=10^{12.5}$~GeV),
  Ref.~\cite{Berezinsky:2002nc} (their Fig.~14 with $n=2.4$,
  $\gamma=2.6$, $E_\mathrm{max}=10^{12}$~GeV), and
  Ref.~\cite{Fodor:2003ph} (their Fig.~3 with $\gamma=2.57$, $n=3.3$,
  $E_\mathrm{max}=3\times10^{12}$~GeV). {\bf Right Panel:} The spectra
  of cosmogenic $\nu_e+\bar\nu_e$ and $\nu_\mu+\bar\nu_\mu$ from the
  CMB background from Ref.~\cite{Engel:2001hd} (their Fig.~4). We have
  normalized our calculations (dashed lines) to match the high energy
  fall-off.}
\label{comparison}
\end{figure*}

Neutron decay contributes then in the Boltzmann equation in the quantities
\begin{equation}
  \Gamma^\mathrm{dec}_n = (\gamma_n\tau_n)^{-1}\quad\mathrm{and}\quad\gamma^\mathrm{dec}_{ni} = \Gamma^\mathrm{dec}_n\,\frac{\mathrm{d}N_i}{\mathrm{d}E}\,.
\end{equation}
for $i=\bar\nu_e,e^{-}$. For neutrons with energy less than
$10^{11}$~GeV the decay length is always smaller than the interaction
length in the photon backgrounds ({\it cf.}~left panel of
Fig.~\ref{scales}). In this case it is convenient to approximate the
production of neutrons as
\begin{align}
  \Gamma^\mathrm{eff}_{pe^-} &= \Gamma^\pi_{pe^-} + \int\mathrm{d}E_n \Gamma^\pi_{pn}\frac{\mathrm{d}N_{e^-}}{\mathrm{d}E}\,,\\
  \Gamma^\mathrm{eff}_{p\bar\nu_e} &= \Gamma^\pi_{p\bar\nu_e} + \int\mathrm{d}E_n \Gamma^\pi_{pn}\frac{\mathrm{d}N_{\bar\nu_e}}{\mathrm{d}E}\,,\\[1ex]
  \Gamma^\mathrm{eff}_{pp} &= \Gamma^\pi_{pp} + \Gamma^\pi_{pn}\,.
\end{align}
We have used this approximation in all our calculations.

\section{Evolution of the Infrared-Optical Background}\label{appendixIR}

The CIB spectrum has been studied and tabulated in
Ref.~\cite{Franceschini:2008tp} for red-shifts up to $z=2$. The
red-shift dependence is given by
\begin{multline}
  n_{\rm CIB}(z,(1+z)E)\\ = (1+z)^2\int_z^\infty\D z'
  \frac{1}{H(z')}\mathcal{L}_{\rm CIB}(z',(1+z')E)\,,
\end{multline}
where $\mathcal{L}_{\rm CIB}$ is the co-moving luminosity density of
the sources and we neglect absorption effects other than expansion. We
assume that this follows the star formation rate: $\mathcal{L}_{\rm
  CIB}(z,E) \propto \mathcal{H}_{\rm SFR}(z)\,\mathcal{L}_{\rm
  CIB}(0,E)$. We can then derive the bolometric evolution as
\begin{equation}
  \frac{N_{\rm CIB}(z)}{N_{\rm CIB}(0)} = (1+z)^3\frac{\int_z^\infty\D z'\,
    \mathcal{H}_{\rm SFR}/(H(z')(1+z'))}{\int_0^\infty\D z'\,
    \mathcal{H}_{\rm SFR}/(H(z')(1+z'))}\,,\label{HIR}
\end{equation}
where $N_{\rm CIB}(z)$ is the number of infrared--optical photons per
proper volume at red-shift $z$. For comparison, the CMB evolves as
$N_{\rm CMB}(z)/N_{\rm CMB}(0) = (1+z)^3$. To simplify the numerical
evaluation we approximate the evolution with redshift as
\begin{equation}\label{NIR}
  n_{\rm CIB}(z,\epsilon) \approx \frac{1}{1+z}\,\frac{N_{\rm CIB}(z)}{N_{\rm CIB}(0)}\, n_{\rm CIB}(0,\epsilon/(1+z))\,.
\end{equation}
The redshift scaling of the quantities $\gamma_{ij}$, $\Gamma_i$,
$b_i$ and $\beta_i$ for the CIB is then obtained from the
corresponding scaling given for the CMB in
Eqs.~(\ref{scaling1}/\ref{scaling2}) and
(\ref{scaling3}/\ref{scaling4}), by multiplying the r.h.s.~by a factor
\mbox{$N_{\rm CIB}(z)/N_{\rm CIB}(0)/(1+z)^3$}. The evolution of the CIB
photon number density is shown in the right panel of
Fig.~\ref{scales}.

\section{Solution of the Boltzmann Equations}\label{appendixBE}

We can express the system of partial integro-differential equations~(\ref{diff0}) in terms of
a system of ordinary integro-differential equations,
\begin{align}\label{diff1}
  \dot{\mathcal{E}}_i &= -H\mathcal{E}_i - b_i(z,\mathcal{E}_i)\,,\\
  \dot Z_i &=
  \big[\beta_i(z,\mathcal{E}_i)-\Gamma_{i}(z,\mathcal{E}_i)\big]\,Z_i+(1+z)\mathcal{L}^\mathrm{eff}_i(z,\mathcal{E}_i)\,,\label{diff2}
\end{align}
where we have defined $\beta_i(z,E) \equiv \partial_Eb_i(z,E)$ and
$Z_i(z,E) \equiv (1+z)Y_i(z,\mathcal{E}_i(z,E))$. The quantity
$\mathcal{E}_i(z,E)$ gives the energy that a particle of type $i$ had
at redshift $z$ if we observe it today with energy $E$ and take into
account CEL. The effective source term in Eq.~(\ref{diff2}) is
\begin{multline}
  \mathcal{L}^\mathrm{eff}_i =\mathcal{L}_i +\sum_j\int\mathrm{d}
  E\,\,\partial_E\mathcal{E}_j\,\gamma_{ji}(z,\mathcal{E}_j,E_i)\,\frac{Z_j}{1+z}\,,\label{Leff}
\end{multline}
where $\mathcal{E}_j(z\,,E)$ and $Z_j(z,E)$ are subject to the
boundary conditions $\mathcal{E}_j(0,E) = E$ and
$Z_j(z_\mathrm{max},E) = 0$. The flux of CRs or neutrinos at $z=0$ can
be expressed as
\begin{widetext}\begin{equation}\label{celflux}
    J_i(E) = \frac{1}{4\pi}Z_i(0,E) =\frac{1}{4\pi}\int_0^\infty\D z\exp\left[\int_0^{z}\D z' \frac{\beta_i(z',\mathcal{E}_i(z',E)) - \Gamma_i(z',\mathcal{E}_i(z',E))}{(1+z')H(z')}\right]
    \frac{1}{H(z)}\mathcal{L}^\mathrm{eff}_i(z,\mathcal{E}_i(z,E))\,.
\end{equation}\end{widetext}

In our calculation we use a logarithmic bin-size of the proton and
neutrino energies with $\Delta\log_{10}E = 0.05$ between $10^5$~GeV
and $10^{15}$~GeV. For the numerical evolution of the differential
equations~(\ref{diff1}) and (\ref{diff2}) we choose a step-size
$\Delta z=10^{-4}$ from 0 to 8. The corresponding step-size in the
propagation distance $\Delta r = c\Delta t$ is then always smaller
than the proton interaction length. In Fig.~\ref{comparison} we
compare our results to previous calculations of cosmic proton and
cosmogenic neutrino spectra, adopting the same values of the Hubble
parameter, source evolution and distribution and the form of the
injection spectrum. The agreement is satisfactory in all cases.

\end{appendix}

\end{document}